\documentstyle[11pt,twoside,paspconf,epsf]{article}

\begin{document}

\title{Dating of H II Regions from Ionization Modeling}

\author{Donald R. Garnett}
\affil{Steward Observatory, University of Arizona, 933 N. Cherry Ave.,
Tucson, AZ  85721, USA}

\begin{abstract}
The emission-line spectrum of an H~II region essentially reflects the 
shape of the EUV spectral energy distribution of the OB stars that ionize 
the gas. In principle one can use this spectrum to extract information 
about the age of the underlying association. In practice, a number of 
effects can lead to systematic errors in derived ages. The star-gas 
geometry is an important consideration, as it determines the ionization 
parameter; high-spatial-resolution imaging is thus essential to constrain
the geometry. Dust grains absorb ionizing photons and decrease the Balmer 
equivalent widths and derived Lyman-continuum luminosity; differential 
extinction between stars and gas can also affect equivalent widths. 
Starburst synthesis models appear to over-predict the numbers of very hot 
Wolf-Rayet stars, especially for metal-rich starbursts; this can have 
deleterious effects on the modeling for such objects. 
\end{abstract}

\section{Introduction}

The motivation for ``dating'' H~II regions is to obtain information about
the distribution of ages in the underlying young stellar population, 
especially in regions where the ionizing stars are difficult to resolve 
from the background population. By doing so one hopes to determine the 
details of the recent star formation history of a region and to understand 
better the typical duration of starburst activity. This provides important 
data for models of the propagation of star formation in starbursts.

Estimating ages for the ionizing stellar population of an H~II region or
starburst is potentially straightforward. However, there are a number of
pitfalls and sources of systematic error that can produce biased results.
I discuss the methodology and some potential difficulties here.

\section{What Do We Really Mean by the ``Age'' of an H II Region?}

When we talk about deriving the ``age'' of an H~II region, what we really 
mean is that we are attempting to derive the EUV spectral energy distribution 
of the ionizing stellar population. One then hopes that the derived SED is
representative of the underlying stellar population. Because of the high 
opacity of the interstellar medium to photons with energies $>$ 13.6 eV, 
the EUV radiation of young OB stars is absorbed by gas in the vicinity; the 
gas becomes ionized and downconverts the EUV photons into lower-energy 
recombination-line and continuum photons, and forbidden line emission from 
electron-impact excitation of ions of elements heavier than helium, which 
escape the H~II region to be observed by us. 

To first order, the strength of the optical/UV forbidden lines relative to 
the recombination lines depends directly on the mean ionizing photon energy; 
the higher the mean photon energy, the higher the electron temperature, 
leading to stronger forbidden-line emission. Thus, as the ionizing cluster 
ages and the hottest stars fade away, the forbidden lines should weaken. One 
well-known complication is that the forbidden lines determine the nebular
cooling, and so the abundance of the heavy elements is the main determinant
of electron temperature. Because the far-IR [O~III] lines are typically the 
most important coolant, a high-metallicity H~II region illuminated by hot O 
stars may have (paradoxically) weaker optical forbidden lines than one ionized
by cooler stars. In addition, the ionization parameter $U$ (a measure of the 
ratio of photon flux density to local gas density) affects the electron 
temperature. The result is that the H~II region modeling problem is highly
degenerate, with the ionizing SED, the ionization parameter, and the heavy 
element abundances all affecting the nebular excitation (Shields 1990). It
is possible to explain the giant H~II region emission-line sequence in a 
galaxy like M101 by either varying the stellar SED at fixed $U$ or by 
varying $U$ at fixed stellar SED (see Figure 1). 

\begin{figure}[t]
\plotfiddle{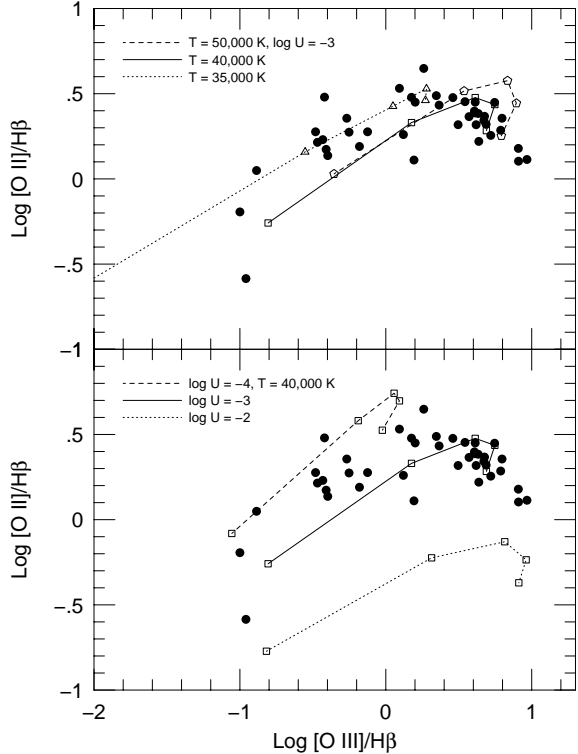}{4.0in}{000}{42}{42}{-125}{-010}
\caption{Spectral sequence for H~II regions in the spiral galaxy M101
(Kennicutt \& Garnett 1996). Overplotted are ionization model sequences 
in which the stellar effective temperature is varied at fixed ionization
parameter ({\it top panel}), and sequences in which the ionization 
parameter is varied at fixed $T_{eff}$ ({\it lower panel}). Nebular 
abundances vary from 0.1 to 2.0 times the solar abundances along each
sequence.} 
\label{fig-1}
\end{figure}

\section{The H~II Region Modeling Process}

Figure 1 of Garcia-Vargas et al. (1997) provides an illustrative flow chart 
outlining the process of ionization modeling of a starburst region.
One starts with a set of observed constraints on physical parameters, which
are used as inputs to a photoionization model constructed with a code such
as CLOUDY. These include: (1) the gas density, determined from a forbidden
line ratio such as [S~II] $\lambda$6717/$\lambda$6731, under the assumption
that the gas is fully ionized; (2) the outer radius $R_{out}$ and the inner
radius $R_{in}$, presumably determined from a high-resolution H$\alpha$
image; (3) the ionizing photon luminosity $N(Ly-c)$ or $Q(H^0)$, derived 
from the H$\alpha$ or H$\beta$ luminosity; and (4) the nebular abundances
determined from the H~II region spectrum if possible. Additional inputs are 
the theoretical spectral energy distribution from the stellar population, 
and the element abundance set if one has not derived them directly from the 
H~II spectrum. One then computes a model and compares the output spectrum 
with the observed spectrum. If they match, the input SED is presumed correct. 
If not, then a new input SED is introduced, and the process iterated until a 
match to the observed spectrum is obtained. A number of hidden assumptions
are often involved: (i) that the stellar IMF is fully sampled. This condition
may be violated for low-luminosity H~II regions with a small number of O stars.
(ii) That there is a single, simple geometry. (iii) That the local recent star 
formation history is of a simple form. 

A number of observable line/ion ratios that are sensitive to the age of the
stellar association are available to constrain the models.

\begin{enumerate}
\item Optical-UV forbidden line/Balmer line ratios: As noted above, the
forbidden-to-Balmer line ratios are sensitive to electron temperature, with
the caveat that they are also very sensitive to metallicity.
\item H$\beta$ equivalent width: The emission equivalent width of H$\beta$
reflects the EUV/optical luminosity ratio. As the age of the stellar 
association increases, EW(H$\beta$) decreases. It is always observed to
be smaller than predicted for a single O star of a given $T_{eff}$ because
evolved massive stars and lower mass stars contribute to the continuum at 
4861 \AA, but not to the ionizing continuum. The predicted EW can be affected 
by uncertainties in the predicted numbers and colors of evolved stars, while 
the observed EW is reduced by the effects of dust grains (see below).
\item He$^+$/H$^+$ ratio: The ratio of ionized He to ionized H is sensitive
to stellar effective temperature for O stars cooler than 40,000 K (Figure
2.5 of Osterbrock 1989). Interpretation of this ratio can be complicated
by uncertainty in the He/H abundance ratio. 
\item $\eta$ parameter: V\'\i chez \& Pagel (1988) proposed using the ion
ratio 
\begin{displaymath}
\eta = { {O^+/O^{+2}}\over {S^+/S^{+2}} }
\end{displaymath}
as a measure of the hardness of the stellar SED for H~II regions. This 
parameter is indeed sensitive to the SED when the ion ratio can be derived
directly from the spectrum. When the electron temperature is unknown and
the ion ratios cannot be derived, an alternative parameter 
\begin{displaymath}
\eta^{\prime} = { {[O II]/[O III]}\over {[S II]/[S III]} }
\end{displaymath}
has been proposed. However, the $\eta^{\prime}$ parameter is much more 
sensitive to metallicity effects (Bresolin, Kennicutt, \& Garnett 1999), 
because each emission line can arise in zones having different electron 
temperatures. 
\end{enumerate}

One can impose additional constraints, such as the colors of the underlying
continuum and the strengths of stellar absorption lines (e.g., Garcia-Vargas
et al. 1997). These are sensitive to uncertainties in the stellar evolution
models, and are beyond the scope of this discussion.

In principle, the problem can be solved and the age of the underlying 
stellar population established, given sufficiently accurate inputs. In
practice, there are a number of complications that can bias age estimates.

\section{Complications}

\subsection{Nebular Geometry}

Nebular geometry is very important for constructing a detailed photoionization
model of an individual H~II region, as it is the principal determinant of 
the ionization parameter. The ionization parameter is determined by the 
combination of the radius parameters $R_{in}$, $n_H$ and $\epsilon$. $R_{out}$
is then determined by the model in the ionization-bounded case, with the 
observed nebular size as a constraint. It is typical to assume a sphere of 
uniform density $n_H$ or a ``ball of filaments'' of clumped gas with some 
$n_H$ and a filling factor $\epsilon$. The latter is a somewhat more realistic 
approximation to a real nebula which has a range of local ionization parameters. 

HST imaging of nearby giant H~II regions resolve these regions into a wealth 
of structure. The typical structure is, in fact, a bright core consisting of 
compact ionized shells around individual OB clusters (presumably evacuated
by the combined action of radiation pressure and stellar winds), with a low 
surface brightness halo of multiple supershells which are probably powered 
by older star formation episodes. Examples can be found at \hfill\break
http://www.oposite.edu/pubinfo/PR/96/27.html \hfill\break
and \hfill\break
http://www.oposite.edu/pubinfo/PR/96/31.html . \hfill\break
The 30 Doradus nebula shows similar structure.

Modeling the high surface brightness shells in these structures by filled 
spheres will result in too high an ionization parameter, because some of 
the gas is too close to the ionizing stars in the filled case. This would
cause one to compensate by using cooler (older) ionizing stars and lead
to too large an age. On the other hand, including the diffuse halo structure
would lead to too extended a structure and a lower ionization parameter.
High resolution imaging should be used whenever possible to constrain the
nebular geometry.

Each H~II region will have its own geometry, which will likely depend on the
local star formation history and the age of the population. Therefore, a
`one-size-fits-all' approach based on a grid of ionization models with a
single geometry is unlikely to obtain correct age estimates for any individual
H~II region or starburst. Ideally, one could build a gas hydrodynamical model
for the evolution of the ionized region as a framework for the ionization
model, but this is probably beyond present capabilities.

\subsection{Dust}

Dust has two major effects on the H~II region spectrum. First, dust grains
mixed with the ionized gas absorb Lyman-continuum radiation. Second, 
obscuration by dust is typically patchy; differential extinction between
stars and gas can affect the emission line equivalent widths.

\begin{figure}[t]
\plotfiddle{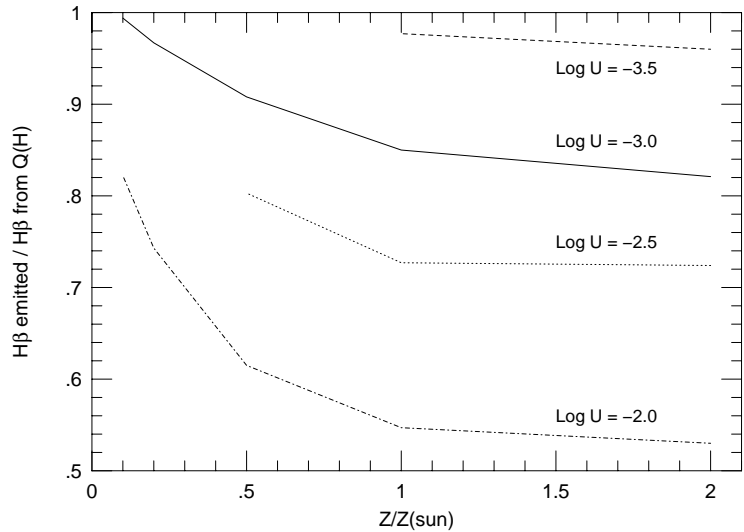}{3.1in}{000}{40}{40}{-150}{000}
\caption{The ratio of emitted H$\beta$ emission to that predicted from the
stellar ionizing photon luminosity as a function of the nebular abundances, 
showing the effects of absorption of ionizing photons by dust grains. The
models are for a stellar temperature of 40,000 K and assume a linear increase 
of the dust-to-gas ratio with metallicity. } 
\label{Garnett-fig2}
\end{figure}

The absorption cross-section for standard interstellar dust grains extends
well into the EUV spectral region with a peak near 17 eV. Dust grains are 
thus quite capable of absorbing ionizing photons in the H~II regions, and
in fact can compete with H and He. When this occurs, the flux of Balmer
line emission is reduced over the dust-free case. Figure 2 displays a set 
of ionization models showing the reduction in H$\beta$ line emission over
that expected from the number of ionizing photons for dusty H~II regions.
I have assumed standard interstellar grains (Martin \& Rouleau 1990), with
a dust-to-gas ratio that varies linearly with metallicity over the range
0.1-2.0 solar O/H. The models show that grains can reduce the emitted H$\beta$
flux by as much as 50\%. The amount lost depends strongly on the ionization
parameter, increasing for higher ionization parameters. A region with high
$U$ is likely to be a young one where the gas is close to the star cluster;
thus more H$\beta$ photons are missed, and EW(H$\beta$) reduced the most,
for the youngest clusters. 

Incidentally, the same phenomenon leads one to underestimate the number
of ionizing photons. Therefore, claims of leakage of ionizing photons
from H~II regions, based on comparing N(Ly-c) from Balmer lines fluxs 
with that estimated from the OB star population, must be viewed with
some skepticism.

Differential extinction between the stars and the gas can also affect 
EW(H$\beta$).
Calzetti et al. (1994) found that the obscuration toward starburst clusters
tended to be lower than that toward the ionized gas. They determined that,
on average, $A_V$ toward the stars was about one-half of that toward the
gas. This is understandable if the stars have evacuated a cavity in the
ionized gas through the combined effects of radiation pressure and stellar
winds. The average derived obscuration for H~II regions in spirals is $A_V$
$\approx$ 1 mag. If $A_V$(stars) is only 0.5 mag, then the observed EW(H$\beta$)
will be about 40\% lower than the intrinsic value. 

\begin{figure}[ht]
\plotfiddle{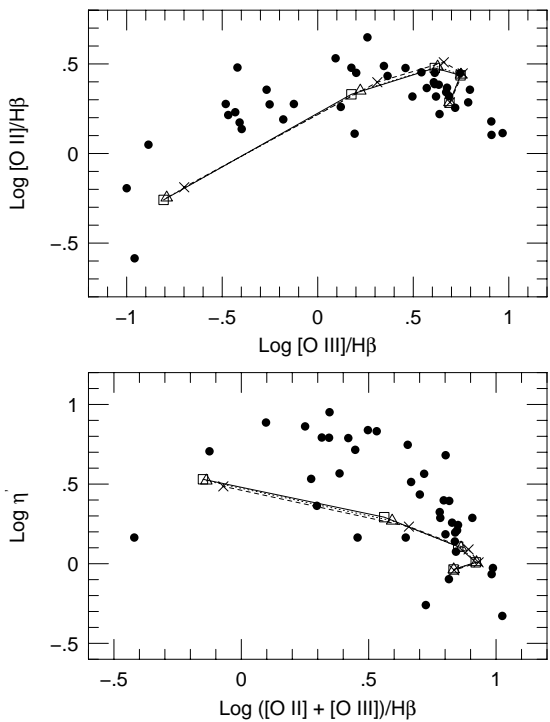}{4.0in}{000}{40}{40}{-140}{000}
\caption{The effects of dust grains on forbidden-line strengths in H~II
regions. Three sequences of models are shown, with $T_{eff}$ = 40,000 K
and log $U$ = $-$3. {\it Solid line plus squares:} dust-free models; {\it dashed line 
+ crosses:} models with standard ISM grains; {\it dotted line plus triangles:} 
models with Orion-type grains. The effects of grains on the emission-line 
ratios are seen to be modest.}
\label{Garnett-fig3}
\end{figure}

These results suggest that dust effects can easily cause one to underestimate
the intrinsic EW(H$\beta$), even for metallicities as low as 0.1 solar O/H.
This would lead to a systematic bias toward larger ages for the stellar 
population. One should therefore exercise caution in weighting EW(H$\beta$) 
as a constraint on the synthesis models.

By contrast, the effects of dust on the relative forbidden-line strengths are
modest (Figure 3), except at high metallicities (Shields \& Kennicutt 
1995). One exception is in the case of very hot ionizing stars, where ionization 
of grains can lead to additional photoelectric heating of the nebula (e.g.,
Ferland 1999). 

\subsection{Wolf-Rayet Stars}

How to deal with Wolf-Rayet stars in ionization models has been a problem
in the past, because of the lack of information on the ionizing luminosities
and shape of the EUV spectra. A great deal of progress has been made in 
modeling W-R atmospheres, but some uncertainties remain in connecting the 
atmosphere calculations to the appropriate stellar evolution state.

The situation regarding very hot W-R stars is a case in point. Spectrum
synthesis models for starburst clusters predict the sudden appearance of
W-R stars at an age of about 3 Myr, accompanied by a sudden jump in the
flux of ionizing photons with E $>$ 54 eV from hot W-R stars (e.g., Schmutz, 
Leitherer, \& Gruenwald 1992). These hot W-R stars are expected to be 
more important in high-metallicity H~II regions and less important at
low metallicities. 

Observations, however, find H~II regions with nebular He~II emission 
preferentially in metal-poor galaxies. Esteban et al. (1993) studied the
central stars of eight W-R ejecta shells in the Milky Way. The characteristic
temperatures for these W-R stars ranged between 30,000 K and 75,000 K,
except for the one WO star which had a nebula exhibiting nebular He~II
emission. Of the six H~II regions in the Local Group with He II emission, 
five are found in metal-poor galaxies, and only one in the Milky Way 
(Pakull 1991, Garnett et al. 1991). Two of these are ionized 
by WO stars, two by rare WN1 stars, one by the X-ray binary LMC X-1, and 
one by an O7 star (which is still a mystery). Except for possibly the WO 
stars, placement of these stars on standard evolutionary tracks is an 
uncertain business, and thus the predictions of nebular He~II emission 
from starburst clusters must also be considered uncertain.

The hot W-R stars in the synthesis models can have a significant effect 
on the nebular spectrum, especially at high metallicities. Ionization
models based on the synthesized OB clusters show a large increase in
nebular excitation at the onset of the W-R phase (Bresolin et al. 1999). 
Metal-rich model H~II region spectra at this stage have much higher 
excitation than observed metal-rich H~II regions in spiral galaxies. 
Since metal-rich H~II regions often show W-R star features (e.g., 
D'Odorico, Rosa, \& Wampler 1983), one may conclude that the synthesis
models overpredict the flux of ionizing photons above 54 eV. 

\section{Discussion}

Ionization modeling has the potential for providing important information
on the spectral energy distributions and ages of the embedded stellar
clusters, if one takes care in considering the effects of geometry and
dust. High-resolution imaging of H~II regions and starbursts is an 
important tool in this process, and should be obtained and employed
whenever possible. The HST archives are very good place to start.
More work is needed on understanding the contribution of Wolf-Rayet
stars on the ionizing continuum of OB clusters, and to understand the
evolutionary status of the hot W-R stars that power nebular He~II 
emission in H~II regions. 

\acknowledgments

DRG thanks the SOC for inviting him to this conference, and acknowledges 
informative discussions with Claus Leitherer, Werner Schmutz, and Daniel 
Schaerer. This research is supported by NASA through LTSA grant NAG5-7734.

\end{document}